\documentclass[aps,12pt,superscriptaddress]{revtex4-1}
\usepackage{latexsym}
\usepackage{graphicx}
\usepackage{amsmath}
\usepackage{amssymb}
\usepackage{verbatim}
\usepackage{braket}
 \usepackage{float}
\usepackage{breqn}
\usepackage{color}
\usepackage{hyperref}
\date{\today}
\newcommand\beq{\begin{equation}}
\newcommand\eeq{\end{equation}}
\newcommand\bea{\begin{eqnarray}}
\newcommand\eea{\end{eqnarray}}

\makeatletter
\let\cat@comma@active\@empty
\makeatother
\begin{document}
\title{An Interplay of Topology and Quantized Geometric Phase for two Different Symmetry-Class Hamiltonians}

\author{Rahul  S}   \author{Ranjith  Kumar  R}  \author{Y   R  Kartik}
\affiliation{Poornaprajna   Institute  of   Scientific  Research,   4,
  Sadashivanagar,  Bangalore-560  080, India.}   \affiliation{Manipal
  Academy  of Higher  Education,  Madhava Nagar,  Manipal  - 57610  4,
  India.}    \author{Amitava   Banerjee}  \affiliation{Department   of
  Physics,  University of  Maryland, College  Park, MD  20742,  USA.}
\author{Sujit    Sarkar}   \affiliation{Poornaprajna    Institute   of
  Scientific Research, 4, Sadashivanagar, Bangalore-560 080, India.}

\begin{abstract}
  \noindent Study of symmetry, topology and geometric phase can reveal
  many  new  and interesting  results  on  the topological  states  of
  matter. Here we  present a completely new and  interesting result of
  symmetry, topology  and quantization  of geometric phase  along with
  the  physical  explanation  for two different symmetry classes. We present a
  detailed  study   of  the  auxiliary   space  for two different symmetry classes 
  of Hamiltonians. We  show explicitly  that the  origin of  the auxiliary
  space inside the curve is only a necessary condition but it is not a
  sufficient condition for the topological state. One of the most interesting results 
  is that same symmetry-class 
  Hamiltonians show different behaviour  in topology and quantized 
  geometric phase. 
  

\end{abstract}
\maketitle
\textbf{Introduction :}\\
 Symmetry plays a significant role in the study of different physical problems. It represents a transformation that leaves the physical system invariant. In quantum mechanics symmetry transformations can be classified as continuous (rotation, translation) and discrete (parity, lattice translations, time reversal). Continuous symmetry transformations give rise to conservation of probabilities and discrete symmetry
 transformations give rise to the quantum numbers. The second revolution of quantum mechanics is topological states of matter \cite{haldane}. 
Symmetry also plays an important role in the study of topological states of
matter\cite{classify-topo}.  \\     Topological      insulators      and
superconductors are  the important topological states  of matter which
can    be   distinguished    based    on    the   symmetry    constraints
\cite{topo.insu,topo.super}. On  the basis  of presence or  absence of
non-spatial symmetries  like time-reversal, particle-hole  and chiral,
one can classify a single-particle Hamiltonian into different symmetry
classes \cite{symm.topo}.  In each  symmetry class one can distinguish
between topological  distinct phases using the  topological invariant.
There are ten distinct symmetry  classes of random matrices, which can
be   interpreted   as    first-quantized   Hamiltonians   of   certain
non-interacting  fermionic systems  \cite{Altland1997}. Among  the ten
symmetry classes for 1D Hamiltonians, only  the AIII, BDI, D, DIII and
CI classes show topological states \cite{Altland1997}. We present basic symmetry 
classes in table \ref{periodictable}. Thus there will
be topological  quantum phase  transition between two  distinct phases
within  a  symmetry class  by  closing  the  gap.   This can  also  be
characterized by  the quantization  of geometric  phase \cite{sarkar2017topological,sarkar2018quantization}.  

Geometric phase  more commonly known as  Berry phase \cite{berry1984},
is a phase  difference acquired by the state when  subjected to cyclic
adiabatic                                                      process
\cite{griffiths1995introduction,bernevig-topo-insu,stanescu2016introduction}. The
geometric  phase  in a  1D  Bloch  band  system  is called  Zak  phase
\cite{zak}.  For  a given  Bloch wave $\psi_k$  with a  quasi-momentum
  $ k$,    reciprocal   lattice   vector
$   G$,   lattice  spacing   $
d$, the  Zak phase  can be expressed  as $$\varphi_{Zak}
=i\int_{-\frac{G}{2}}^{\frac{G}{2}}\left\langle
u_k|\partial_k|u_k\right\rangle,  $$ where  $u_k(x)=e^{-ikx}\psi_k(x)$
and $G=\frac{2\pi}{d}$.   The physics of geometric  phase reveals many
important aspects of topological state of matter \cite{wilczek1989}.

There have been many breakthroughs in the field of topological quantum
condensed   matter  starting   with   integer   quantum  Hall   effect
\cite{klitzing1980}    and    fractional     quantum    Hall    effect
\cite{tsui1982},  and   later  the  idea  of   topological  insulators
\cite{PhysRevLett.95.146802,    RevModPhys.82.3045,    moore2010birth,
  nishimori2010elements,    PhysRevB.80.155131,    PhysRevB.85.075125,
  PhysRevB.83.035107}.  One of the classic examples of this kind is 1D
topological   superconductor    \cite{topo.super}.   In    this   case
topologically trivial  and non-trivial  phases are distinquished  by 
gap closing.  This can  be characterized by  the Pfaffian  of Majorana
representation  matrix. In  general,  the Pfaffian  $Pf[A]$  of a  $2n
\times 2n$ skew-symmetric matrix A is defined as
\begin{equation}
Pf[A]=       \frac{1}{2^n       n!}\sum_{\sigma\in\Pi_{2n}}       sign
\Pi^{n}_{i=1}A_{\sigma(2i-1),\sigma(2i)}.
\end{equation}
If  $A$ is  a  $2  \times 2$  matrix,  $  A=\left( \begin{matrix}  0&&
  a\\ -a&& 0
\end{matrix}\right), $ then $Pf[A]=a$. In the case of topological states
of matter Pfaffian of Majorana  representation matrix is a topological
invariant. On  the other hand, one  can also write the  Hamiltonian in
the form  of Bogoliubov de  Gennes (BdG) mean-field  Hamiltonian using
Nambu spinor. Then  the anti-unitary particle-hole  constraint of the
BdG  Hamiltonian gives  rise to  the quantization  of geometric  phase
\cite{hatsugai2006quantized}, which indicates  the topological quantum
phase  transition.  The  topological configuration  space of  a system
gives rise to  the particular value of  a topological invariant quantity
like winding number. The closed  curves in the configuration space are
the auxiliary  space curves which  also specify the topological states of
the system. Auxiliary  space curves have a unique  way of representing
the topological quantum  phase transition.  When the system  is in the
topological state,  the auxiliary  space curve encircles  the origin;
for  the non-topological  state,  the origin  lies  outside the  space
curve; and  at the point of  phase transition, the origin  lies on the
space curve in  which case the topological invariant  number is ill
defined \cite{sarkar2018quantization}. 
In the present study we consider two different symmetry classes
of model Hamiltonians. One is BDI and the other is AIII symmetry class. 
These two different classes belong to the ten symmetry classes of the 
topological state of matter (see table \ref{periodictable}).\\
\begin{table}
\begin{center}
\includegraphics[scale=0.2]{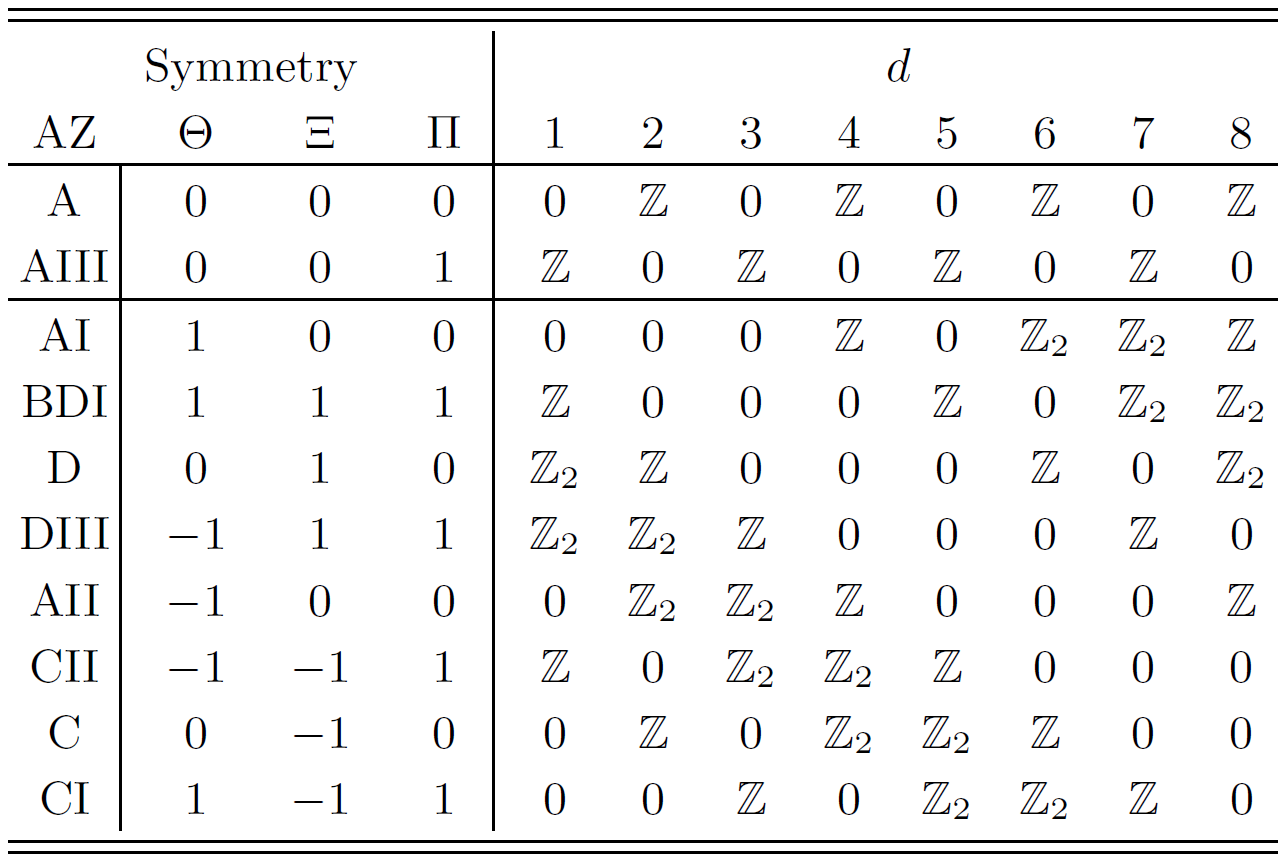}
\end{center}
\caption{Periodic table of topological insulators and topological superconductors \cite{Altland1997}. Here $\Theta$ are time-reversal, $\Xi$ are particle-hole and $\Pi$ are chiral symmetry operators.}
\label{periodictable}
\end{table}
Topological  state of  quantum matter is  a very  rapidly growing
field. One may quantum-simulate different topological properties in different 
  physical systems
  \cite{RevModPhys.86.153,sarkar2015quantum,sarkar2015existence}.   
  Therefore  the different topological symmetry-classes Hamiltonians we present  here may
  be important in the study of quantum simulation physics.
  Simulating  quantum  systems like  topological  state  of matter  is
  practically not possible using classical systems.
  One can  also use quantum devices  that mimic the evolution  of other
  quantum  systems as  quantum simulators  \cite{georgescu2014quantum}.
  Therefore the experimental realization of these quantum devices helps
  us in understanding many interesting quantum systems.

\textbf{Motivation and Goals}:\\
1) We consider four model Hamiltonians ($\mathcal{H}^{(1)}(k)$,$\mathcal{H}^ {(2)}(k)$, $\mathcal{H}^ {(3)}(k)$,$\mathcal{H}^ {(4)}(k)$) in the momentum space and study explicitly 
the symmetry classes of these model Hamiltonians
\cite{0034-4885-78-10-106001,0953-8984-25-36-362201,newrashba}.   \\
2) We study the quantization of geometric phase and also calculate the Pfaffian for different model Hamiltonians to 
characterize the topological phase transition.\\
3) We also study the auxiliary space  curve and its closseness condition for different model Hamiltonians to show explicitly the possibility for topological non-trivial phase to exist. This is also verified by the study of dispersion plots for different model Hamiltonians.\\
4) These new and important results based on two different topological symmetry classes motivate quantum simulation physicists to quantum-simulate these types of model Hamiltonian systems.\\

 { \bf \large Model Hamiltonians:}\\
 \textbf{First case }:We consider the Kitaev's chain in the matrix form as our model Hamiltonian,
 \begin{equation}  \mathcal{H}^{(1)}(k)   =\left(  \begin{matrix}
           0&(\epsilon_k-\mu)    +    2i\Delta   \sin{k} \\ (\epsilon_k-\mu) - 2i\Delta \sin{k} &0
   \end{matrix}\right),\end{equation}
where $\epsilon_k=-2t\cos k$.

 \textbf{Second case}:  Here  we  add variant term, $\alpha k$ to 
  $\sigma_y$  component. This  is also  a plausible  system to  quantum-simulate,  
  since the Hamiltonian  is linear  in momentum
  $k$. With this  motivation we consider the
  Kitaev's     Hamiltonian     with    this     variant term,
  {   \begin{equation}  \mathcal{H}^{(2)}(k)   =\left(  \begin{matrix}
          0&(\epsilon_k-\mu)    +    2i\Delta   \sin{k}    +    i\alpha
          k\\ (\epsilon_k-\mu) - 2i\Delta \sin{k} - i\alpha k&0
  \end{matrix}\right).\end{equation} }
  We will show that this Hamiltonian has the same symmetry properties of Kitaev model and both the Hamiltonians fall under BDI symmetry class.
  
  \textbf{Third case}:  Here  we   consider  the  variant term  in  the
    $\sigma_x$ component. Kitaev's  Hamiltonian  in presence  of
    variant term       becomes        {     \begin{equation}
        \mathcal{H}^{(3)}(k)=\left( \begin{matrix}  0 & (\epsilon_k-\mu  + \alpha
            k)  +  2i\Delta  \sin{k} \\ (\epsilon_k-\mu  +  \alpha  k)  -
            2i\Delta \sin{k} & 0
    \end{matrix}\right).\end{equation} }
    We will show that this Hamiltonian falls under the AIII symmetry class.
    
 \textbf{Fourth  case}:  Here  we   consider  the  variant term  in  both
 $\sigma_x$     and     $\sigma_y$      components,     which     gives
        { \begin{equation} \mathcal{H}^{(4)}(k) =\left( \begin{matrix}
                0&(\xi_k+\alpha_1  k) +  2i\Delta  \sin{k} +  i\alpha_2
                k\\ (\xi_k+\alpha_1  k) - 2i\Delta \sin{k}  - i\alpha_2
                k&0
  \end{matrix}\right),\end{equation} }
        where $\xi_k=\epsilon_k-\mu$. We will show that this Hamiltonian has the same symmetry properties as $\mathcal{H}^{(3)}(k)$  and falls under the AIII symmetry class.
        All these Hamiltonians are hermitian in character.\\
 { \bf \large Results and discussion for the Hamiltonians of BDI symmetry class:}
In this section we present the detailed study and results of the two model Hamiltonians $\mathcal{H}^{(1)}(k)$ and $\mathcal{H}^{(2)}(k)$. $\mathcal{H}^{(1)}(k)$ in the matrix form can be written as
 \begin{equation}
 \mathcal{H}^{(1)}(k)=    \left(   \begin{matrix}    -2t\cos(k)-   \mu    &&
   2i\Delta\sin(k) \\-2i\Delta\sin(k) && 2t\cos(k)+ \mu 
 	\end{matrix}\right). \label{kitaev}
 \end{equation}
Now we study the basic symmetries, i.e.,  time-reversal
 ($\mathcal{T}$),  particle-hole   ($  \mathcal{C}$)  and   chiral  ($
 \mathcal{S}$) symmetry operations for the model Hamiltonian. Among these three basic symmetries time-reversal operator commutes with the Hamiltonian while particle-hole and chiral operators anti-commute with the Hamiltonian \cite{symm.topo,stanescu2016introduction}. The operator forms of these symmetry operations are $\mathcal{T}=K$ (where $K$ is complex conjugate operator), $  \mathcal{C}=\sigma_x K$ (where $\sigma_x$ is one of the Pauli's matrices) and $\mathcal{S}=\sigma_x$. The commutation relations can be checked as\\
 
 \begin{equation}
 \begin{aligned}
 	\mathcal{T}\mathcal{H}^{(1)}(k)\mathcal{T}^{-1}&={K} \mathcal{H}^{(1)}(k) {K}^{-1}   \\                             
 	&= \hat{K} \left( \begin{matrix}
 	-2t\cos(k)- \mu && 2i\Delta\sin(k)\\
 	-2i\Delta\sin(k) && 2t\cos(k)+ \mu \\
 	\end{matrix}\right)  \hat{K}    \\                          
 	&= \mathcal{H}^{(1)}(k),
 \end{aligned}
 \end{equation}
 
 \begin{equation}
 \begin{aligned}
 	\mathcal{C}\mathcal{H}^{(1)}(k)\mathcal{C}^{-1} &= (\sigma_x{K})\mathcal{H}^{(1)}(k)(\sigma_x {K})^{-1}\\
 	&=\left(\begin{matrix}
 	0 && 1\\
 	1 &&  0 \\
 	\end{matrix}\right) \left(\begin{matrix}
 	-2t\cos(k)-\mu  && i2\Delta\sin(k)\\
 	-i2\Delta\sin(k) && 2t\cos(k)+\mu\\
 	\end{matrix}\right) \left(\begin{matrix}
 	0 && 1\\
 	1 &&  0 \\
 	\end{matrix}\right)\\
 	&= -\mathcal{H}^{(1)}(k),
 \end{aligned}
 \end{equation}
 
  \begin{equation}
  \begin{aligned}
  	\mathcal{S}\mathcal{H}^{(1)}(k)\mathcal{S}^{-1} &= (\sigma_x)\mathcal{H}^{(1)}(k)(\sigma_x)^{-1}\\
  	&=\left(\begin{matrix}
  	0 && 1\\
  	1 &&  0 \\
  	\end{matrix}\right) \left(\begin{matrix}
  	-2t\cos(k)-\mu  && i2\Delta\sin(k)\\
  	-i2\Delta\sin(k) && 2t\cos(k)+\mu\\
  	\end{matrix}\right) \left(\begin{matrix}
  	0 && 1\\
  	1 &&  0 \\
  	\end{matrix}\right)\\
  	&= -\mathcal{H}^{(1)}(k).
  \end{aligned}
  \end{equation}
Thus we have
 	\begin{equation*}
          \mathcal{T}\mathcal{H}^{(1)}(k)\mathcal{T}^{-1} = \mathcal{H}^{(1)}(k) ,
          \;\;\;\;   \mathcal    {C}\mathcal{H}^{(1)}(k)\mathcal{C}^{-1}   =
          -\mathcal{H}^{(1)}(k),
 	\end{equation*} 
 	\begin{equation} \mathcal{S}\mathcal{H}^{(1)}(k)\mathcal{S}^{-1} = -\mathcal{H
          }(k).
 \end{equation}
 The  Hamiltonian falls under the  symmetry class BDI of  the ten
 symmetry  classes  of   topological  insulators  and  superconductors
 \cite{Altland1997}  with topological  invariant number  $\mathbb{Z}$,
 which takes integer  values (see table \ref{periodictable}). Each value of  $\mathbb{Z}$ indicates a
 set  of  $\mathcal{H}^{(1)}(k)$  which  can  be  interpolated  continuously
 without  breaking  the  symmetries  and without  closing  the  energy
 gap. Topological quantum  phase transition can be  observed by tuning
 the parameters of $\mathcal{H}^{(1)}(k)$ to get gapless state. This closing
 the gap  involves changes  in $\mathbb{Z}$  by one  unit. To  get the
 condition for the parameters  which distinguish between topological
 trivial and non-trivial phases one  can calculate the Majorana number
 ($\mathcal{M}$).\\
 The  Kitaev's   chain  in lattice is
   \cite{kitaev2001}
   	\begin{dmath}
   		H_0  = [\sum_{n}-{t}  ({c_n}^{\dagger}  c_{n+1} +  h.c
                    )-{\mu}  {c_n}^{\dagger}  {c_n}+{|\Delta|}  (  {c_n}
                    {c_{n+1}}  +  h.c  )  ],  \label{kitaev1}\end{dmath}
          where  $  t$ is  the  hopping  matrix  element, $\mu$  is  the
          chemical potential,  and $|\Delta |$  is the magnitude  of the
          superconducting gap. We write  the Hamiltonian in the momentum
          space as
   
   \begin{dmath} 	H_1  =   \sum_{k> 0} ( \mu + 2t \cos k)
   	({\psi_k}^{\dagger}    {\psi_k}     +    {\psi_{-k}}^{\dagger}
     {\psi_{-k}}) + 2i \Delta \sum_{k  > 0} \sin k ({\psi_k}^{\dagger}
     {\psi_{-k}}^{\dagger} + {\psi_{k}} {\psi_{-k}}),\end{dmath}
   \noindent  where $  {\psi^{\dagger}}_{k}  (\psi_{k})$  is the  creation
   (annihilation) operator of the spinless fermion of momentum $k$.\\
 One can rewrite  eq. \ref{kitaev1} in the  Majorana fermion operators
 by   using   the   relation   $\gamma_{2i-1}=c_i^{\dagger}+c_i$   and
 $\gamma_{2i}=i(c_i^{\dagger}-c_i)$,  where $\gamma_j$  represent real
 fermions    with    properties   $\gamma_j^{\dagger}=\gamma_j$    and
 $\left\lbrace  \gamma_i,\gamma_j \right\rbrace  =2\delta_{ij}$.  Thus
 eq. \ref{kitaev1}  can be written  in the Majorana  representation as
 \cite{kitaev2001}
 \begin{equation}
   \mathcal{H}^{(1)}(k)=\frac{i}{2}         \sum_{\alpha,\beta}        \gamma_{\alpha}
   A_{\gamma\beta} \gamma_{\beta},
   \label{majo-kit}
 \end{equation}
 where $A $ is a real and anti-symmetric matrix.
 
 One can verify  the existence of topological  trivial and non-trivial
 phases by calculating the Pfaffian of Majorana representation matrix,
 with a real orthogonal transformation, $W$, and using the property of
 Pfaffian, i.e.  $Pf[WAW^T]=Pf[A]\det[W]$. 
 One  can write the Majorana
 number $\mathcal{M}=\det(W(0))\det(W(\pi))$.  The  property of $W(k)$,
 i.e. $W(k)^*=W(-k)$, implies quantized geometric phase indicating
 topological quantum phase transition \cite{budich2013equivalent}.
 One can  write the  Hamiltonian in block-diagonal  form by  using the
 real orthogonal transformation,
 \begin{equation}
{\small A(k)=  \left( \begin{matrix}  0 &&  -2t\cos k-\mu+i2\Delta\sin
    k\\ 2t\cos k+\mu+i2\Delta\sin k && 0\\
\end{matrix}\right). }
 \end{equation}
 Majorana number can  be expressed in terms of Pfaffian  of the matrix
 $A(k)$ as
 \begin{equation}
 \mathcal{M}= sign \left\lbrace Pf[A(0)] Pf[A(\pi)]\right\rbrace. \label{pfian}
 \end{equation}
 Pfaffian of the matrix $A(k)$ at $k=0$ and $k=\pi$ is calculated as
 \begin{equation}
 Pf[A(0)]= -(2t+\mu), \;\;\;\; Pf[A(\pi)]=2t-\mu.
 \end{equation}
 For  the  parametric  condition  $\mu  >  2t$,  the  Majorana  number
 $\mathcal{M}=+1$ indicates that the  system is in the non-topological
 phase;  also,  for  $\mu<2t$, the  Majorana  number  $\mathcal{M}=-1$
 indicates that  the system is in  the topological phase. It  is clear
 from   this  that   the  topological   phase  transition   occurs  at
 $\mu=\pm2t$.
 
\begin{figure}
 	\includegraphics[width=6cm,height=8cm]{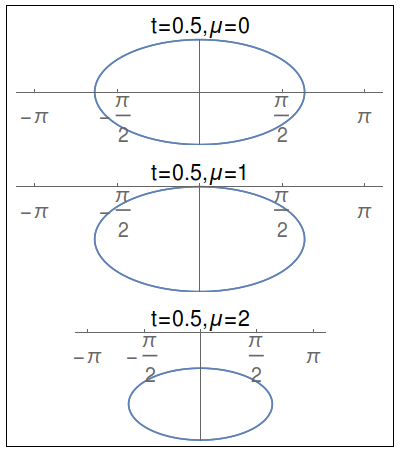}
 	\caption{Parametric plots for $\mathcal{H}^{(1)}(k)$ Hamiltonian for different
          values of $\mu$. } \label{aux1}
 \end{figure}

 Existence  of  topological states  can  also  be confirmed  from 
 fig. \ref{aux1}.
  This  shows the  auxiliary space  of the  Kitaev chain,  which is  a
  closed trajectory with  origin inside the curve.  The integral along
  this trajectory  takes quantized  values indicating  the topological
  quantum phase transition.  The value of Zak phase is  $\pi$ when the
  origin is inside the  curve of the auxiliary space and  0 when it is
  outside.  Since $\mathcal{H}^{(1)}(k)$  has anti-unitary  particle-hole symmetry  it
  gives  the  reality  condition  for  Majorana  representaion  matrix
  ($A(k)^*=A(-k)$), i.e.
  \begin{equation}
  \mathcal{C}A(k)\mathcal{C}^{-1} = A(-k).
  \end{equation}
  This results in the quantized Zak phase which is $0$ for trivial and
  $\pi$  for non-trivial  phases. We  verify this  by calculating  the
  geometric phase  for the  system. First we  write Kitaev's  chain in
  momentum space as
$$ H_{k}= (\epsilon_{k}- \mu) \sigma_{z} - 2\Delta\sin k \sigma_{y}.$$
  One can also write the same Hamiltonian in a rotated basis as
  \begin{dmath} H_{k} =  (\epsilon_{k}- \mu) \sigma_{x} - 2\Delta\sin k \sigma_{y},
  \end{dmath}
 where $\epsilon_{k} = -2t\cos k$.
\begin{dmath}   H_{k} = \begin{bmatrix}
 0 && (\epsilon_{k}- \mu)\\ (\epsilon_{k}- \mu) && 0 \\
 \end{bmatrix}   - 2\Delta\sin k \begin{bmatrix}
 0 && -i\\
 i && 0\\
 \end{bmatrix}, \end{dmath}

 \begin{dmath} H_{k} = \begin{bmatrix}
 0  &&  (\epsilon_{k}-   \mu)+2i\Delta\sin  k\\  (\epsilon_{k}-  \mu)-
 2i\Delta\sin k && 0\\
 \end{bmatrix}, \end{dmath}

 \noindent  where  $  \epsilon_{k}  -   \mu  =  r\cos\theta  $  and  $
 2\Delta\sin k = r\sin\theta.$ Now the Hamiltonian is reduced to
 $$ H= \begin{bmatrix}
 0  && r e^{i\theta} \\
 r e^{-i\theta} && 0 
 \end{bmatrix}, $$ 
 
 \noindent where $  r = \sqrt{(2t\cos k+\mu)^2 +  4\Delta^2 \sin^2 k}$
 and   \;\;$    \theta   =    -\tan^{-1}(\frac{2\Delta\sin   k}{2t\cos
   k+\mu}).$

 Here we  are calculating  the geometric phase  for the  lowest-energy
 eigenfunction.   The  basic  aspects of  the  geometric  phase  are
 given in the appendix.  The Berry  phase is given by $ \gamma_{n}
 =  i   \oint  \left\langle  \Psi_{n}|\nabla_{R}|\Psi_{n}\right\rangle
 dR,$

 where\;\;  $  \Psi_{n}  =  \left(  \frac{-1}{\sqrt{2}}\right)  \left(
 -e^{i\tan^{-1}(\frac{2\Delta\sin   k}{\mu+2t\cos  k})}\right),$   and
 finally the Berry connection is given by
 $$  A(k)=\left\langle  \Psi_{n}|\dfrac{d\Psi_{n}}{dk}\right\rangle  =
 \dfrac{\Delta(2t+\mu\cos  k)}{(\mu+2t\cos k)^2+4\Delta^2\sin^2  k}.$$
 Finally,
 \begin{equation} \gamma_{n} =  i \oint A dk = i\oint_C \; \;
   \dfrac{\Delta(2t+\mu\cos(k))}{(\mu+2t\cos(k))^2+4\Delta^2\sin^2(k)}dk .
 \end{equation}
 
  \begin{figure}[H]
    \centering \includegraphics[scale=0.34]{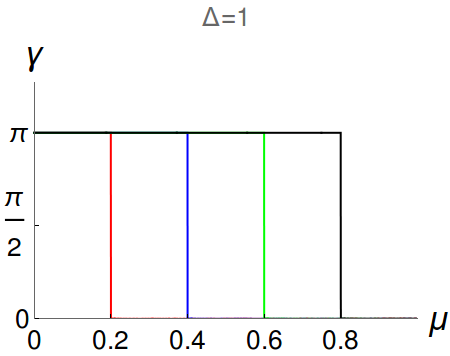}
    
  	\caption{(color online) Variation of  $\gamma$ with $\mu$. The
          red, blue, green,  and black curves are for $t$  = 0.1, 0.2,
          0.3 and 0.4 respectively.} \label{gamma.mu}
  \end{figure}
\begin{figure}
	\includegraphics[width=7cm,height=6cm]{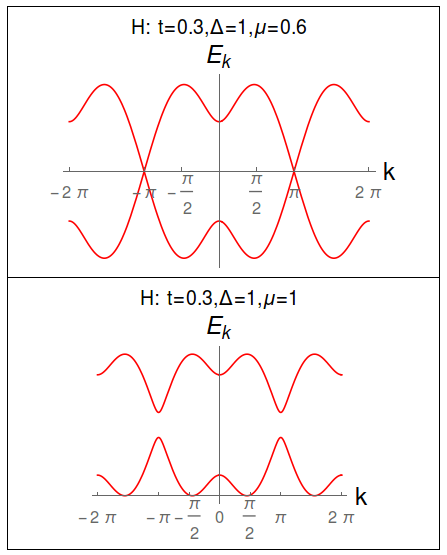}
	\caption{Dispersion  curve $E_{k}$  vs k  for the  Hamiltonian
          $\mathcal{H}^{(1)}(k)$.} \label{Ek}
\end{figure}

In fig.  \ref{gamma.mu},  we present the variation  of geometric phase
($\gamma$) with  $\mu$. It is  clear from this  study that there  is a
topological quantum  phase transition from $\gamma=\pi$  to $\gamma=0$
\cite{zaktopo}. We  have also observed  that the transition  occurs at
$\mu=2t$ \cite{Kaisun}.  \\ The physical explanation of  this transition
can be  understood in the  following way: Fig. \ref{Ek}  describes the
energy dispersion spectrum of this model Hamiltonian, where we observe
the gapless state at $k = \pm  \pi$. A gapless state is present if the
value of the  parameters obey the transition relation $\mu  =\pm 2 t$.
This   is   when   transition   takes  place   from   topological   to
non-topological state.   If the values  of the parameters do  not obey
the  transition  relation,  then  we   observe  a  gapped  state  (non-topological state), 
as shown in the lower panel of fig. \ref{Ek}.

Here  we clearly  observe  that  the gapless  states,  in other  words
degenerate states, appear  for discrete values of $\mu$,  i.e.  $\mu =
\pm 2t$  for $k=0$  and $k=\pm\pi$,  not for  the different  values of
$\mu$.  Thus we justify the topological characterization of the Kitaev
chain from  the perspective  of symmetry Pfaffian  number calculation,
winding  number  calculation  and  also   the  gapless  state  at  the
topological quantum phase transition point.

\begin{figure}[h!]
		\includegraphics[width=5.5cm,height=9.5cm]{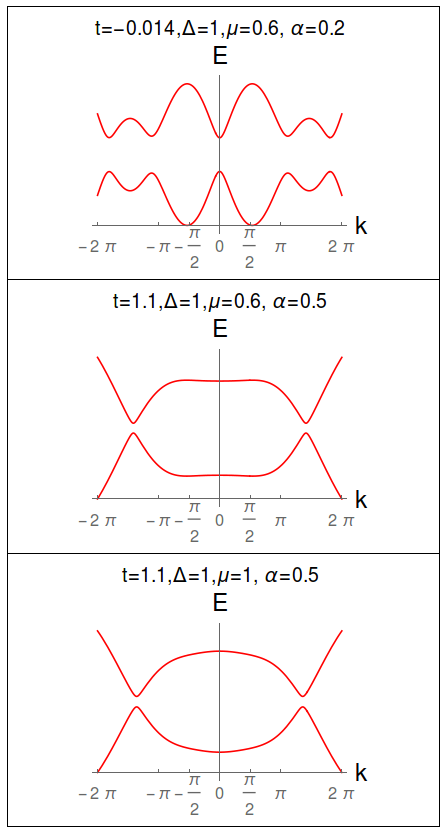}
		\caption{Dispersion  curve of  $E_{k}$  vs  k for  the
                  $\mathcal{H}^{(2)}(k)$ Hamiltonian.} \label{Ek2}
\end{figure}

\textbf{ Results and discussion for the Hamiltonian $\mathcal{H}^{(2)}(k)$ :}
The Hamiltonian $\mathcal{H}^{(2)}(k)$ in the matrix form can be written as
\begin{equation}
	 \mathcal{H}^{(2)}(k)     =\left(\begin{matrix}    -2t\cos(k)-\mu     &&
           2i\Delta\sin(k)+i\alpha  k\\ -2i\Delta\sin(k)-i\alpha  k &&
           2t\cos(k)+\mu
	 \end{matrix} \right).
\end{equation}
 Here we consider three variants of Kitaev chain in momentum space. One can express the Kitaev chain in the basis of $\sigma_x$ and $\sigma_y$. The present study is only for the theoretical exercise over the Kitaev chain. But it has some interesting physics that all these variants of Kitaev chain ($\mathcal{H}^ {(2)}(k), \mathcal{H}^ {(3)}(k),\mathcal{H}^ {(4)}(k)$) are hermitian  in character. Upon adding of the $\alpha k$ term, either in $\sigma_x$ or $\sigma_y$,  the two components of the Kitaev chain, we observe very interesting results from the perspective of different topological symmetry classes of the auxiliary space and the condition for topological characterization. The pairing mechanism of the Kitaev chain in real space is the well-defined p-wave pairing mechanism. But for the other three model Hamiltonians ($\mathcal{H}^ {(2)}(k), \mathcal{H}^ {(3)}(k),\mathcal{H}^ {(4)}(k)$), we do not know the nature of the interaction. We consider these model Hamiltonians for the complete theoretical interest and at the same time they satisfy the hermiticity property of the Hamiltonian and also one can express these model Hamiltonians into two different symmetry classes (BDI and AIII). Kitaev's model belongs to the BDI symmetry class. Therefore the present study has relevence from the perspective of symmetry class. We would like to study how the symmetry and topological properties are modified in the presence of additional terms. Therefore we present an extensive study from the perspective of topological symmetry class in momentum space. \\
 Many studies in theoretical physics do not find proper physical justification at the very beginning. But in due course of time, they are justified. The new and important results of this study may motivate quantum simulation physicists to quantum-simulate this system.\\
This  model Hamiltonian satisfies the  conditions for  time-reversal, particle-hole  and
chiral symmetry operations, 
 \begin{equation}
 \begin{aligned}
 	 \mathcal{T}\mathcal{H}^{(2)}(k)\mathcal{T}^{-1} &=  {K}\mathcal{H}^{(2)}(k) {K}^{-1}\\
 	 &= \hat{K} \left(  \begin{matrix}
 	 -2t\cos(k)-\mu && 2i\Delta\sin(k)+i\alpha k\\
 	 -2i\Delta\sin(k)-i\alpha k && 2t\cos(k)+\mu\\
 	 \end{matrix}\right) \hat{K}\\
 	 &= \mathcal{H}^{(2)}(k),
 \end{aligned}
 \end{equation}
 	 	 		 
 \begin{equation}
 \begin{aligned}
 	 \mathcal{C}\mathcal{H}^{(2)}(k)\mathcal{C}^{-1} &= (\sigma_x{K})\mathcal{H}^{(2)}(k)(\sigma_x {K})^{-1}\\
 	 &= \left(\begin{matrix}
 	 0 && 1\\
 	 1 &&  0 \\
 	 \end{matrix}\right) \left(\begin{matrix}
 	 -2t\cos(k)-\mu  && 2i\Delta\sin(k)+i\alpha k\\
 	 -2i\Delta\sin(k)-i\alpha k && 2t\cos(k)+\mu\\
 	 \end{matrix}\right) \left(\begin{matrix}
 	 0 && 1\\
 	 1 &&  0 \\
 	 \end{matrix}\right)\\
 	 & = -\mathcal{H}^{(2)}(k),
 \end{aligned}
 \end{equation}
 	 	 	
 \begin{equation}
 \begin{aligned}
 	 \mathcal{S}\mathcal{H}^{(2)}(k)\mathcal{S}^{-1} &= \sigma_x \mathcal{H}^{(2)}(k)\sigma_x^{-1} \\
 	 &= \left(\begin{matrix}
 	 0 && 1\\
 	 1 &&  0 \\
 	 \end{matrix}\right) \left(\begin{matrix}
 	 -2t\cos(k)-\mu  && 2i\Delta\sin(k)+i\alpha k\\
 	 -2i\Delta\sin(k)-i\alpha k && 2t\cos(k)+\mu\\
 	 \end{matrix}\right) \left(\begin{matrix}
 	 0 && 1\\ 
 	 1 &&  0 \\
 	 \end{matrix}\right)\\
 	 &= -\mathcal{H}^{(2)}(k).
 \end{aligned}
 \end{equation}
 Thus we have
 \begin{equation*}
  \mathcal{T}\mathcal{H}^{(2)}(k)\mathcal{T}^{-1}=\mathcal{H}^{(2)}(k),
  \;\;\;    \mathcal{C}\mathcal{H}^{(2)}(k)\mathcal{C}^{-1}     =    -
  \mathcal{H}^{(2)}(k),
 \end{equation*}
 \begin{equation}
 \mathcal{S}\mathcal{H}^{(2)}(k)\mathcal{S}^{-1}=                    -
 \mathcal{H}^{(2)}(k).
 \end{equation}
As in the case of Kitaev  chain ($\mathcal{H}^{(1)}(k)$), $\mathcal{H}^{(2)}(k)$ also belongs to
the  symmetry  class  BDI   \cite{Altland1997}  with  the  topological
invariant $\mathbb{Z}$ (see table \ref{periodictable}).   One can also expect  the topological quantum
phase  transition  in  $\mathcal{H}^{(2)}(k)$.   This  system  has  anti-unitary
particle-hole symmetry,  thus one may  consider that the system is  in the
topological    state    and    Zak    phase    must    be    quantized
\cite{sarkar2017topological,zak}. But we observe  that even though the
symmetry  class of  the Hamiltonian  $\mathcal{H}^{(1)}(k)$ and  $\mathcal{H}^{(2)}(k)$ are  the
same,   there   is  no   topological   non-trivial   phase  for   these model Hamiltonians. 
Addition of the term $\alpha k$ breaks the anti-symmetric property of
the Majorana representation matrix for $k=\pi$,
\begin{equation}
 {\small A(\pi)= \left( \begin{matrix}
 0 && 2t-\mu+i \alpha \pi\\
 -2t+\mu+i \alpha \pi && 0\\
 \end{matrix}\right) .}
  \end{equation}
  For  this case  one cannot  calculate the  Majorana number  since  the Pfaffian  
  does  not  exist  due  to  lack  of the
  anti-symmetric  property.   This also  indicates  that  there is  no
  topological non-trivial phase  for this system. This  result is also
  evident from the nature of auxiliary space of $\mathcal{H}^{(2)}(k)$.
  \begin{figure}[h!]
  \includegraphics[width=14cm,height=5cm]{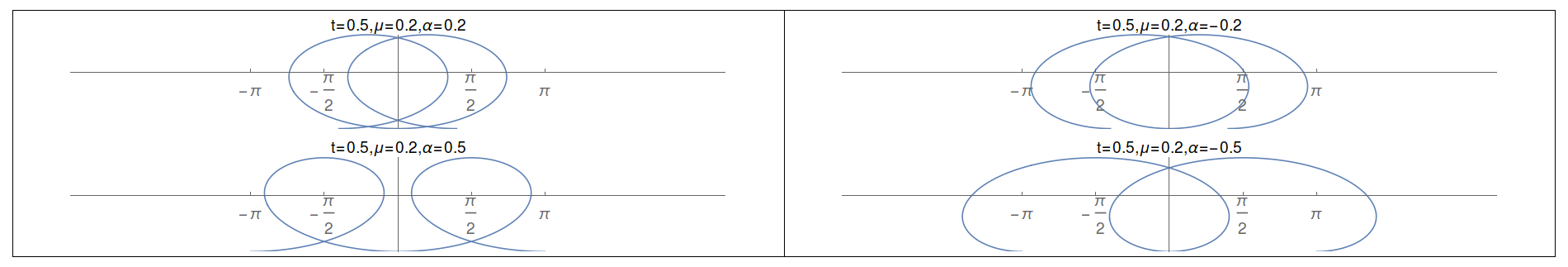}
  \caption{Parametric plots  for $\mathcal{H}^{(2)}(k)$ for different  values of
    $\mu$, $\alpha$ and $t$. }\label{aux3}
  \end{figure}
  Fig. \ref{aux3} shows that the  trajectory in the auxiliary space is
  not  closed and  the integral  along  the trajectory  will not  take
  quantized  values.   We observe  an  interesting  feature from  this
  behaviour in the  auxiliary space that, although the  origin of auxiliary
  space   is  encircled   by  the   curve,  the   system  is   in  the
  non-topological state. At the same  time there is no mirror symmetry
  with $\alpha$.  This can also be verified from the energy dispersion
  curve (fig.   \ref{Ek2}).  It  shows there is  no gapless  state for
  topological quantum phase  transition to occur.  
  The  Hamiltonian  $\mathcal{H}^{(2)}(k)$ does  not have  the
  mirror-symmetric  auxiliary space  curves for positive  and negative
  values of $\alpha$.\\
  
  It is clear from this section that even though both the Hamiltonians $\mathcal{H}^{(1)}(k)$ and $\mathcal{H}^{(2)}(k)$ belong to the BDI symmetry class they have different topological properties. In the case of $\mathcal{H}^{(2)}(k)$ there is no topological non-trivail phase as expected for the BDI symmetry class. Majorana representation matrix is anti-symmetric for $\mathcal{H}^{(1)}(k)$ while it does not satisfy this condition for $\mathcal{H}^{(2)}(k)$. Thus it is not possible to calculate topological invariant to characterize the topological state for $\mathcal{H}^{(2)}(k)$. Energy dispersion for $\mathcal{H}^{(1)}(k)$ is gapless, while for $\mathcal{H}^{(2)}(k)$ there is no gapless dispersion, indicating there is no topological phase transition for $\mathcal{H}^{(2)}(k)$. Auxiliary space curve also gives the same conclusion. It is closed curve for $\mathcal{H}^{(1)}(k)$, indicating integer topological invariant. In the case of $\mathcal{H}^{(2)}(k)$ it is open curve, which shows the absence of topological non-tivial phase.
  Thus it is clear from this study that the same symmetry class does not reveal the same topological properties.\\
  
\begin{figure}[H]
	\centering \includegraphics[width=8.8cm,height=8cm]{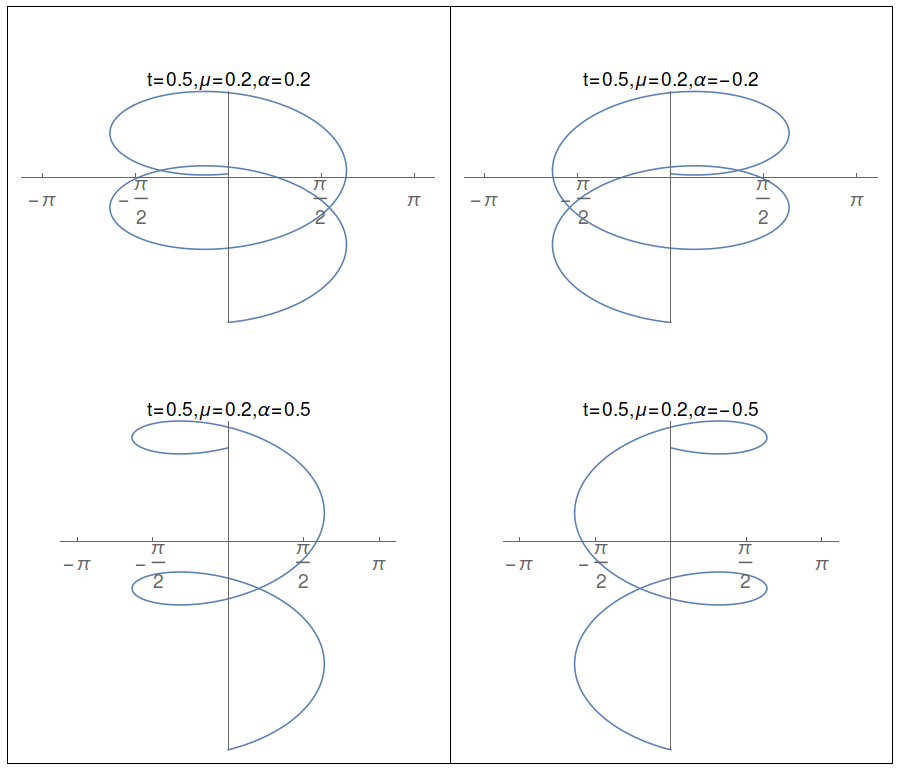}
\caption{Parametric  plots for  $\mathcal{H}^{(3)}(k)$ for  different values  of
  $\mu$, $\alpha$ and $t$.}\label{aux2}
\end{figure}

 \textbf{\large Results and discussion for the Hamiltonians of AIII symmetry class}   \\   
Here we present the results for Hamiltonians $\mathcal{H}^{(3)}(k)$ and $\mathcal{H}^{(4)}(k)$.
 The matrix form of the Hamiltonian $\mathcal{H}^{(3)}(k)$ can be written as
 \begin{equation}
 \mathcal{H}^{(3)}(k)  =   \left(  \begin{matrix}  -2t\cos(k)-\mu-\alpha   k  &&
   2i\Delta\sin(k)\\ -2i\Delta\sin(k) && 2t\cos(k)+\mu+\alpha k\\
 	\end{matrix}\right).
 \end{equation} 
 We observe that  $\mathcal{H}^{(3)}(k)$ does not satisfy  the condition for
 time-reversal  and particle-hole  symmetry  operations but  satisfies
 only the chiral symmetry condition.
 
 \begin{equation}
 \begin{aligned}
 	 \mathcal{T}\mathcal{H}^{(3)}(k)\mathcal{T}^{-1} &=  {K} \mathcal{H}^{(3)}(k)  {K}^{-1}\\
 	 &= \hat{K} \left(  \begin{matrix}
 	 -2t\cos(k)-\mu-\alpha k && 2i\Delta\sin(k)\\
 	 -2i\Delta\sin(k) && 2t\cos(k)+\mu+\alpha k\\
 	 \end{matrix}\right) \hat{K}\\
 	 &\neq\mathcal{H}^{(3)}(k),
 \end{aligned}
 \end{equation}

 \begin{equation}
 \begin{aligned}
 	 \mathcal{C}\mathcal{H}^{(3)}(k)\mathcal{C}^{-1} &= (\sigma_x {K}) \mathcal{H}^{(3)}(k) (\sigma_x {K})^{-1}\\
 	  &=   \left( \begin{matrix}
 	  0 && 1\\
 	  1 && 0\\
 	  \end{matrix}\right)  \left( \begin{matrix}
 	  -2t\cos(k)-\mu+\alpha k && 2i\Delta\sin(k)\\
 	  -2i\Delta\sin(k) && 2t\cos(k)+\mu-\alpha k\\
 	  \end{matrix}\right)  \left( \begin{matrix}
 	  0 && 1\\
 	  1 && 0\\
 	  \end{matrix}\right)\\
 	 &\neq - \mathcal{H}^{(3)}(k),
 \end{aligned}
 \end{equation}
 
 \begin{equation}
 \begin{aligned}
 	\mathcal{S}\mathcal{H}^{(3)}(k)\mathcal{S}^{-1} &= (\sigma_x) \mathcal{H}^{(3)}(k) (\sigma_x)^{-1}\\
 	 &= \left( \begin{matrix}
 	 0 && 1\\
 	 1 && 0\\
 	 \end{matrix}\right) \left(  \begin{matrix}
 	 -2t\cos(k)-\mu-\alpha k && i2\Delta\sin(k)\\
 	 -i2\Delta\sin(k) && 2t\cos(k)+\mu+\alpha k\\
 	 \end{matrix}\right)  \left( \begin{matrix}
 	 0 && 1\\
 	 1 && 0\\
 	 \end{matrix}\right) \\
 	 &= - \mathcal{H}^{(3)}(k).
 \end{aligned}
 \end{equation}
Thus we have
 \begin{equation*}
  \mathcal{T}\mathcal{H}^{(3)}(k)\mathcal{T}^{-1}\neq\mathcal{H}^{(3)}(k)
  \;\;\;   \mathcal{C}\mathcal{H}^{(3)}(k)\mathcal{C}^{-1}    \neq   -
  \mathcal{H}^{(3)}(k),
 \end{equation*}
 \begin{equation}
 \mathcal{S}\mathcal\mathcal{H}^{(3)}(k)\mathcal{S}^{-1}=                    -
 \mathcal\mathcal{H}^{(3)}(k).
 \end{equation}
 Thus, from the symmetry analysis,  $\mathcal{H}^{(3)}(k)$ falls under AIII class
 in  the  ten  symmetry  classes  with  topological  invariant  number
 $\mathbb{Z}$.  This  indicates that  there is  a possibility  for the
 topological  quantum phase  transition  by tuning  the parameters  to
 obtain the gapless state with the change in the topological invariant
 number $\mathbb{Z}$. But we observe a strange behaviour of the system,
 that  it  does  not   allow  one  to  calculate  Majorana  number
 $\mathcal{M}$ (eq. \ref{pfian}) to get  the condition for the  parameters. The
 Majorana representation matrix  $A_{\alpha\beta}$ for $\mathcal{H}^{(3)}(k)$ is
 given by
 \begin{equation}
 {\small A(k)= \left( \begin{matrix}  0 && \xi_k+\alpha k+i2\Delta\sin
     k\\ -\xi_k+\alpha k+i2\Delta\sin k && 0\\
 \end{matrix}\right) ,}
  \end{equation}
where  $\xi_k=-2t\cos k-\mu$.   Here $A(k)$  loses the  anti-symmetric
property  for $k=\pi$,  which  does  not allow  one  to calculate  the
Pfaffian  of the  matrix. This  indicates  that there  is no  Majorana
number,  which implies  that the  system is  in non-trivial  topological
phase. This can also be verifed by the trajectory of the system in the
auxiliary space, given in fig. \ref{aux2}.
A  very interesting  observation can  be made  from  fig. \ref{aux2}:
although the  auxiliary space curve  in the upper panel  encircles the
origin,  the  curve is  not  closed.   For the  system  to  be in  the
topological state, the auxiliary space curve encirling the origin is a
necessary condition, but the closedness  of the curve is the sufficient
condition, as we  observe in the Kitaev chain  (fig.  \ref{aux1}). The
left and  the right  panels represent the  auxiliary space  curves for
positive and negative values of $\alpha$. They show mirror-symmetric   
behaviour   for   positive    and   negative   values   of
$\alpha$. Although the auxiliary space curve is symmetric with respect
to positive  and negative values  of $\alpha$, it is  not symmetric
with respect to $k$. Thus we conclude that the topological properties of
the system do not depend on the symmetry of the auxiliary space.
\begin{figure}[H]
	\centering \includegraphics[width=6.5cm,height=8cm]{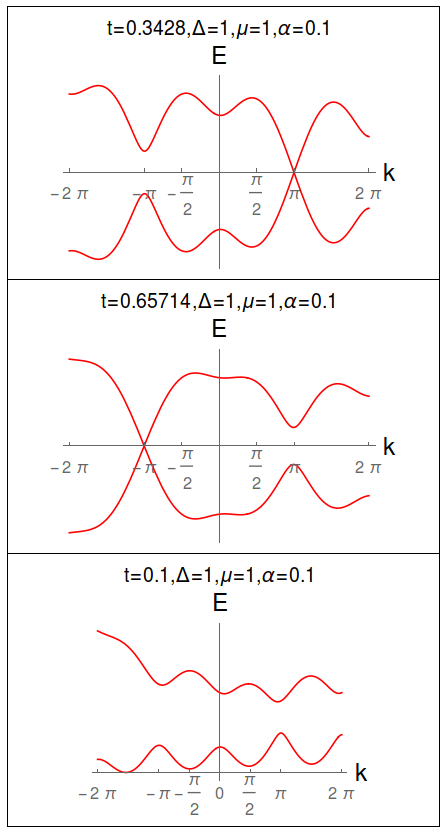}
	\caption{Dispersion curve of $E_{k}$  vs k for the Hamiltonian
          $\mathcal{H}^{(3)}(k)$.} \label{Ek1}
\end{figure}
\noindent  We  observe that  there  is  no  closed trajectory  in  the
auxiliary space. We  also  observe that  the  addition  of  $\alpha  k$  to
Hamiltonian $\mathcal{H}^{(3)}(k)$ results in breaking of the periodicity of
the Brillouin zone. One can  observe this from energy dispersion curve
(fig. \ref{Ek1}) that $E(\pi) \neq  E(-\pi)$. This lack of periodicity
does   not   allow   one   to  calculate   the   geometric/Zak   phase
\cite{sarkar2017topological,zak}. In other words the integral over the
non-periodic Brillouin zone will not be a Cauchy integral and does not
take  the  quantized  value  \cite{ablowitz2003complex},  which  again
indicates that there is no topological quantum phase transition. We discuss this in the next section. 

\noindent  \textbf{Results  and  discussion  for  the Hamiltonian $\mathcal{H}^{(4)}(k)$ :} The  matrix form  of the Hamiltonian  $\mathcal{H}^{(4)}(k)$ is
given by
\begin{equation}
	 \mathcal{H}^{(4)}(k)=\left(                    \begin{matrix}
           -2t\cos(k)-\mu-\alpha_1   k  &&   2i\Delta\sin(k)+i\alpha_2
           k\\ -2i\Delta\sin(k)-i\alpha_2  k && 2t\cos(k)+\mu+\alpha_1
           k \\
	 \end{matrix} \right).
\end{equation}
We  observe that  $\mathcal{H}^{(4)}(k)$  belongs to  the  symmetry class  AIII,
i.e. it only obeys the chiral symmetry condition.

\begin{equation}
\begin{aligned}
	 \mathcal{T}\mathcal{H}^{(4)}(k)\mathcal{T}^{-1} &=  {K} \mathcal{H}^{(4)}(k) {K}^{-1}\\
	 &= \hat{K} \left(  \begin{matrix}
	 -2t\cos(k)-\mu-\beta_1 k && 2i\Delta\sin(k)+i\beta_2 k\\
	 -2i\Delta\sin(k)-i\beta_2 k && 2t\cos(k)+\mu+\beta_1 k \\
	 \end{matrix}\right) \hat{K}\\
	 &\neq \mathcal{H}^{(4)}(k),
	 \end{aligned}
	 \end{equation}

\begin{equation}
\begin{aligned}
     \mathcal{C}\mathcal{H}^{(4)}(k)\mathcal{C}^{-1} &= (\sigma_x \hat{K}) \mathcal{H}^{(4)}(k) (\sigma_x {K})^{-1}\\
	 &=   \left( \begin{matrix}
	 0 && 1\\
	 1 && 0\\
	 \end{matrix}\right)  \left( \begin{matrix}
	 -2t\cos(k)-\mu+\beta_1 k && 2i\Delta\sin(k)+i\beta_2 k\\
	 -2i\Delta\sin(k)-i\beta_2 k && 2t\cos(k)+\mu-\beta_1 k \\
	 \end{matrix}\right) \left(  \begin{matrix}
	 0 && 1\\
	 1 && 0\\
	 \end{matrix}\right)\\
	 &\neq - \mathcal{H}^{(4)}(k),
	 \end{aligned}
	 \end{equation}

\begin{equation}
\begin{aligned}
	 \mathcal{S}\mathcal{H}^{(4)}(k)\mathcal{S}^{-1} &= (\sigma_x)\mathcal{H}^{(4)}(k) (\sigma_x)^{-1}.\\
	 &= \left( \begin{matrix}
	 0 && 1\\
	 1 && 0\\
	 \end{matrix}\right) \left(  \begin{matrix}
	 -2t\cos(k)-\mu-\beta_1 k && 2i\Delta\sin(k)+i\beta_2 k\\
	 -2i\Delta\sin(k)-i\beta_2 k && 2t\cos(k)+\mu+\beta_1 k \\
	 \end{matrix}\right)  \left( \begin{matrix}
	 0 && 1\\
	 1 && 0\\
	 \end{matrix}\right)\\
	 &= - \mathcal{H}^{(4)}(k).
\end{aligned}
\end{equation}
Thus we have
\begin{equation*}
  \mathcal{T}\mathcal{H}^{(4)}(k)\mathcal{T}^{-1}\neq\mathcal{H}^{(4)}(k),
  \;\;\;   \mathcal{C}\mathcal{H}^{(4)}(k)\mathcal{C}^{-1}    \neq   -
  \mathcal{H}^{(4)}(k),
 \end{equation*}
 \begin{equation}
 \mathcal{S}\mathcal{H}^{(4)}(k)\mathcal{S}^{-1}=                    -
 \mathcal{H}^{(4)}(k).
 \end{equation}
  \begin{figure}[h!]
  \includegraphics[width=8.5cm,height=12cm]{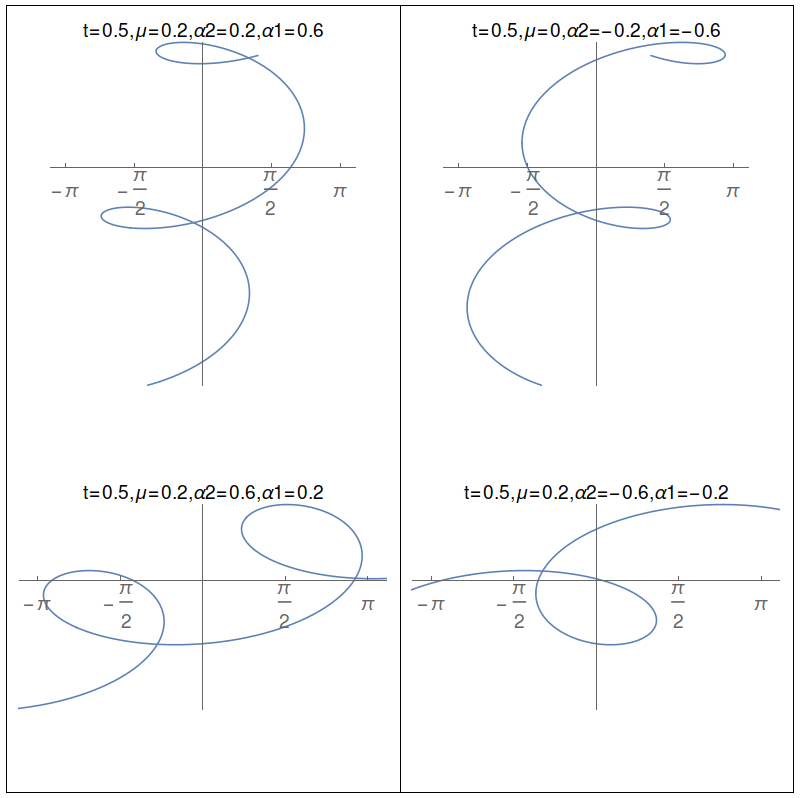}
   \caption{Parametric plots for $\mathcal{H}^{(4)}(k)$  for different values of
     $\mu$, $\alpha$ and $t$.}\label{aux4}
  \end{figure}
 Thus Hamiltonian $\mathcal{H}^{(4)}(k)$ also falls under the same symmetry class as $\mathcal{H}^{(3)}(k)$ i.e. AIII symmetry class. Here  also  one can  expect  the
 topological  quantum phase  transition with  change in  the value  of
 topological   invariant   number   $\mathbb{Z}$.   But   similar   to
 $\mathcal{H}^{(2)}(k)$ and $\mathcal{H}^{(3)}(k)$, we observe the Majorana representation
 matrix breaks its  anti-symmetric property as we  add the $\alpha k$
 term,
\begin{equation}
 {\small  A(\pi)= \left(  \begin{matrix}  0  && 2t-\mu+\alpha_1  \pi+i
     \alpha_2 \pi\\ -2t+\mu+\alpha_1 \pi+i \alpha_2 \pi && 0\\
 \end{matrix}\right) .}
  \end{equation}
 Thus Pfaffian does not exist for this system, which shows there is no
 topological non-trivial phase.
 \begin{figure}[h!]
   		\includegraphics[width=8cm,height=10cm]{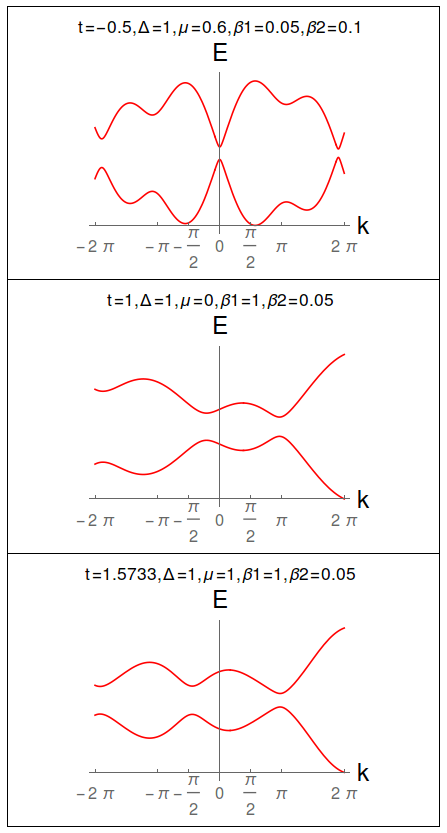}
   		\caption{Dispersion  curve of  $E_{k}$  vs  k for  the
                  $H^{(4)}(k)$ Hamiltonian.} \label{Ek3}
   \end{figure}
 Here the trajectory in the auxiliary space is not closed and integral
 along the trajectory  is not quantized. Absence of  the origin inside
 the perimeter of the trajectory in  fig. \ref{aux4} shows there is no
 topological state for the system.  The curves in the auxiliary space
 are  neither symmetric  with  $\alpha$ nor  with $k$.  A  curve in  the
 auxiliary space  encircling the origin  is a necessary  condition but
 the closedness of the curve is the sufficient condition for the system
 to  be in  the  topological state.  This can  also  be verified  from
 fig.   \ref{Ek3},    which   shows   the   energy    dispersion   for
 $\mathcal{H}^{(4)}(k)$. We observe that there are  no gapless states responsible for
 topological  quantum phase transition.\\
 
 It is clear from this section that both the Hamiltonians $\mathcal{H}^{(3)}(k)$ and $\mathcal{H}^{(4)}(k)$ belong to the symmetry class AIII. They are expected to get the topological non-trivial phase for this symmetry class but both the Hamiltonians do not show any property indicating the presence of topological non-trivial phase. Majorana representaion matrix does not show anti-symmetric property, and auxiliary space curve is not closed and thus shows there is no possibility of having topological non-trivial phase. \\
 
 \noindent \textbf{A Comparision of Results for two different symmetry classes :}\\
 We have noticed in the BDI symmetry class that the topological behaviour of the Hamiltonians $\mathcal{H}^{(1)}(k)$ and $\mathcal{H}^{(2)}(k)$ are different. The Kitaev Hamiltonian $\mathcal{H}^{(1)}(k)$ shows the topological phase transition while there is no topological phase transition for $\mathcal{H}^{(2)}(k)$. It results from the dispersion curve, which for $\mathcal{H}^{(1)}(k)$ shows a gapless state indicating the transition point. But for BDI symmetry-class Hamiltonians obey $\mathcal{H}(k=-\pi)=\mathcal{H}(k=\pi)$, while for the case of AIII symmetry $\mathcal{H}(k=-\pi)\neq\mathcal{H}(k=\pi)$. We also observe that there is no closing of the gap at both $k=\pm \pi$, as we find for $\mathcal{H}^{(1)}(k)$ dispersion. The energy dispersion for $\mathcal{H}^{(3)}(k)$ and $\mathcal{H}^{(4)}(k)$ are different although these two belong to the same symmetry class. Thus it is clear from this study of two different symmetry classes that gap closedness is the prime condition to get topological quantum phase transition of the system. \\
 
 \noindent\textbf{  \large A   Few  Relevant  Calculations  and   Discussion  for  the
   Topological   Characterization   of   these   two symmetry classes  - BDI and AIII.}

 \textbf{(A).    A derivation for sufficient    condition     for   topological
   characterization from the behaviour of curves in auxiliary space.}

 Here  we  mathematically  prove  the  sufficient  condition  for  the
 topological  characterization of  the  system from  the behaviour  of
 auxiliary space.  We have
   \begin{equation}
   \mathcal{H}^{(3)}(k)=(-2t\cos    k    -    \mu   +    \alpha_1k)\sigma_x    +
   (\alpha_2k-2\Delta\sin k)\sigma_y.
   \end{equation}
   We plot the parametric plot $(x(k),y(k))$,
   \begin{equation}
   x(k)=-2t\cos k-\mu+\alpha_1 k = r(k)\cos \theta(k),
   \end{equation}
   \begin{equation}
   y(k)=\alpha_2k-2\Delta\sin k = r(k)\sin \theta(k),
   \end{equation}
   so that,  in the auxiliary plane,  \begin{equation} r^2(k)=(-2t\cos
     k-\mu+\alpha_1 k)^2 + (\alpha_2k-2\Delta\sin k)^2,
   \end{equation}
   \begin{equation}
   \theta(k)=  \tan^{-1}\left[ \frac{\alpha_2k-2\Delta\sin  k}{-2t\cos
       k-\mu+\alpha_1 k} \right] .
   \end{equation}
   To  have   a  closed   curve  for   $k$  running   between  $\left[
     -\pi,\pi\right] $ the curve must come back to its starting point,
   i.e.
   \begin{equation}
   r^2(k=\pi)=r^2(k=-\pi),  \;\;\;\;  \theta(k=\pi)=\theta(k=-\pi)\;\;
   mod(2\pi).
   \end{equation}
   Putting  these two  conditions in  the expression  of $r^2(k)$  and
   $\theta(k)$,
   \begin{dmath}
   	(-2t\cos        (-\pi)-\mu+\alpha_1         (-\pi))^2        +
     (\alpha_2(-\pi)-2\Delta\sin       (-\pi))^2      =       (-2t\cos
     (\pi)-\mu+\alpha_1 (\pi))^2 + (\alpha_2(\pi)-2\Delta\sin (\pi))^2
     ,\label{r}
   \end{dmath}
   and
   \begin{dmath}
   	\tan^{-1}\left[               \frac{\alpha_2(-\pi)-2\Delta\sin
            (-\pi)}{-2t\cos   (-\pi)-\mu+\alpha_1    (-\pi)}   \right]
        =\tan^{-1}\left[               \frac{\alpha_2(\pi)-2\Delta\sin
            (\pi)}{-2t\cos          (\pi)-\mu+\alpha_1          (\pi)}
          \right]. \label{theta}
   \end{dmath}
   The conditions  for the curve to  be closed, i.e. eq.   \ref{r} and
   eq.     \ref{theta},   can    be   simultaneously    satisfied   if
   $\alpha_1=\alpha_2=0$.  Thus it is clear  from this study that this
   equation  ($\alpha_1=  0  =\alpha_2)$  is  only  satisfied  by  the
   Hamiltonian $\mathcal{H}^{(1)}(k)$, i.e. Kitaev chain.\\
   Therefore it is clear from this study that the addition of $\alpha k$ term either in $\sigma_x$ or $\sigma_y$ components or both does not favour the existence of the topological state of matter. The field of quantum simulation physics is rich and it has application in different branches of physics \cite{sarkar2017topological,sarkar2015quantum}. Our strong belief is that this result motivates quantum simulation physicists to find this behaviour in different physical systems. To the best of our knowledge this
   study is  the first in  the literature  to study the  necessary and
   sufficient conditions for  topological characterization from the
   behaviour of auxiliary space.\\
   
    \textbf{(B). A  general physical explanation for  the existence of
      topological state: from the  perspective of Berry connection and
      geometric phase.  }
    
    BdG  Hamiltonian  obeys  an anti-unitary  particle-hole  symmetry,
    $\mathcal{C}=\sigma_xK$ ($\sigma_x$ and $K$  are Pauli spin matrix
    and complex conjugate operator respectively.),
   \begin{equation}
   \left\lbrace \mathcal{H}_{BdG}, \mathcal{C}\right\rbrace ,
   \end{equation}
   with $\mathcal{C}^2=1$. This symmetry  implies that the bands below
   and above the energy gap are conjugates of each other, i.e.
   \begin{equation}
   \mathcal{C}\ket{\Psi^o(-k)}=e^{-i\phi(k)} \ket{\Psi^e(-k)}.
   \end{equation}
   Here $\ket{\Psi^o}$ and $\ket{\Psi^e}$ are  the Bloch states of the
   occupied  and empty  bands respectively.  Using this  abelian Berry
   connection \cite{hong1996berry,pachos2006geometric} of the occupied
   bands can be written as
   \begin{equation}
   \mathcal{A}^o(k)                        =                        -i
   \sum_{\alpha}\bra{\psi^o_{\alpha}(k)}\partial_k\ket{\psi^o_{\alpha}(k)},
   \end{equation}
   where $\alpha  = 1, .   .  .   ,n$ represent the  independent Bloch
   bands. One can  observe the equivalence of the  Berry connection of
   the occupied bands at $k$ and the empty bands at $-k$ up to a gauge
   transformation,
   \begin{equation}
   \mathcal{A}^o(-k)  =   \mathcal{A}^e(k)-  \sum_{\alpha}  \partial_k
   \phi_{\alpha}(k).
   \label{gaug}
   \end{equation} 
   This  constraint  implies that  the  Zak  phase  over half  of  the
   Brillouin zone can be written as
   \begin{equation}
   \gamma=\int\limits_{-\pi}^{\pi}\mathcal{A}^o(k)         dk        =
   \int\limits_{0}^{\pi}\left[        \mathcal{A}(k)-       \partial_k
     \phi(k)\right] dk, \label{geo.phase}
   \end{equation}
    with $  \mathcal{A}(k)=\mathcal{A}^o(k)+\mathcal{A}^e(k)$.  In the
    Majorana representation  one can  write the $\mathcal{H}_{BdG}$  in diagonal
    form as
    \begin{equation}
    W(k)\mathcal{H}_{BdG}W^{\dagger}(k)=\sigma_z diag(\epsilon_{\lambda}),
    \end{equation}
    where from the particle-hole symmetry $W$ safisfies the condition
    \begin{equation}
    \mathcal{C}W(k)\mathcal{C}^{-1}=W(-k).
    \end{equation}
    This    implies   that    the    phase    factor   $\phi(k)$    in
    eq. \ref{geo.phase} vanishes. Thus in the presence of anti-unitary
    particle-hole  symmetry  the Zak  phase  is  quantized to  integer
    multiples of  $\pi$, indicating the gapless  state with topological
    quantum phase transition.
    
    But for the  present case, in presence of  $\alpha k$ term, one cannot
    express  the Berry  connection  $\mathcal{A}^{0}(-k)$ to  $\mathcal{A}^{e}(k)$ by  
    eq. \ref{gaug}   and  also   geometric  phase,   $\gamma$,  as   in
    eq. \ref{geo.phase}.
    
    \noindent\textbf{C.    Topological   characterization   from   the
      perspective of winding number}
    
In general BdG Hamiltonian in the symmetry class BDI can be written in
the special form as
\begin{equation}
\mathcal{H}^{(1)}(k)=\left(  \begin{matrix}  h_0(k)  &&  i\Delta(k)\\  -i\Delta(k)  &&
  -h_0(k)
\end{matrix}\right).
\end{equation}
This can be written in the diagonal form as
\begin{equation}
\mathcal{H}^{(1)}(k)=\left( \begin{matrix}
0 && A(k)\\
A^T(-k) && 0
\end{matrix}\right) ,
\end{equation}
where   $A(k)=  h_0(k)   +  i\Delta(k)$,   satisfying  the   condition
$A^*(k)=A(-k)$. Winding  number in this  case can be defined  from the
phase of $A(k)$,
\begin{equation}
A(k)=|A(k)|e^{i\theta(k)}.
\end{equation}
Considering $z(k)=e^{i\theta(k)}$ one can define the winding number as
\begin{equation}
w=\frac{-i}{\pi}\int\limits_{0}^{\pi}\frac{dz(k)}{z(k)}              =
\frac{1}{\pi}(\theta(\pi)-\theta(0)).\label{winding}
\end{equation}
In the case of Kitaev chain, i.e. $\mathcal{H}^{(1)}(k)$, we have
\begin{equation}
|A(k)|=\sqrt{(2t\cos k+\mu)^2+(2\Delta\sin k)^2},
\end{equation}
\begin{equation}
\theta(k)=\tan^{-1}\left[ \frac{2\Delta\sin k}{-2t\cos k-\mu}\right].
\end{equation}
Thus the winding number for Kitaev chain is given by
\begin{equation}
w=\frac{1}{\pi}(\pi-0)=1.
\end{equation}
Since $\tan^{-1}(0)$  can be either  $0$ or $\pi$, winding  number can
take quantized values, $w=\pm1,0$.

In the case of $\mathcal{H}^{(2)}(k)$,  $\mathcal{H}^{(3)}(k)$ and $\mathcal{H}^{(4)}(k)$ one cannot
define the  integral in eq.   \ref{winding}.  Even  if we  try to
calculate the winding by brute force for $\mathcal{H}^{(4)}(k)$, then we have
\begin{equation}
\theta(k)=\tan^{-1}\left[   \frac{2\Delta\sin  k+\alpha_2   k}{-2t\cos
    k-\mu+\alpha_1 k}\right].
\end{equation}
The winding number is given by
\begin{equation}
\begin{aligned}
w&=\frac{1}{\pi}\left(          \tan^{-1}\left(         \frac{\alpha_2
  \pi}{2t-\mu+\alpha_1            k}\right)             -            0
\right)\\         &=\frac{1}{\pi}\tan^{-1}\left(        \frac{\alpha_2
  \pi}{2t-\mu+\alpha_1 k}\right).
\end{aligned}
\end{equation}
We observe that the winding number  is not quantized to integer values
but  takes   continuous  values,  thus   the  system  will   be  in
non-topological  state  for  all  non-zero values  of  $\alpha_1$  and
$\alpha_2$.

{ \bf \large Conclusions:}

   We   have  presented   the  results   of  symmetry,   topology  and
   quantization of geometric phase along with the physical explanation
   for two different topological symmetry classes. The nature of  these model Hamiltonians and
   the  results should motivate  quantum simulation  physics to simulate these types of model Hamiltonian in different physical systems. We  have shown
   explicitly  that  the  symmetry criteria are not sufficient to characterize the topological state of
   the systems. We have also presented the results based on
   auxiliary space  to derive the necessary  and sufficient conditions
   for topological characterization.\\
   



\textbf{\large Acknowledgments }\\
The authors would  like to acknowledge DST
(EMR/2017/000898) for the funding and Raman Research Institute library for the books and
journals.  The  authors would like to  acknowledge Mr.N.A.Prakash for critical reading
of  this manuscript.  Finally  authors would like
to                         acknowledge                        International Centre for Theoretical Sciences
Lectures/seminars/workshops /conferences/discussion    meetings   on
different aspects of physics.\\

    
    \bibliography{firstpaper}

    \;\;\;\textbf{\large Appendix} 
    
     \textbf{\large Berry phase in Bloch band}  
     
    The basic Hamiltonian for the Bloch  band can be
    written as $H=\frac{k^2}{2m}+V(r)$,  with the value $V(r+a)=V(r)$,
    where  $a$ is  the distance  between two  lattice points.  The Bloch
    state  satisfies  the following  conditions  for  the edge  state:
    $\psi_{nK}(r+a)=e^{iq.a}\psi_{nk}(r)$,  i.e.  the wavefunction  at
    the points $r$ and $r+a$ are related  to the phase.  One can also write
    the model Hamiltonian as $H(q)=\frac{(k+\hbar q)^2}{2m}+V(r),$ and the
    wavefunction  $$\psi_{nq}(r)=u_{nq}(r)e^{iq.r},$$  where $u_{nq}(r)$  is
    the  free  particle wavefunction  in  the presence  of  periodic
    potential,     which     also      satisfies     the     following
    condition:$$u_{nq}(r+a)=u_{nq}(r).$$ The most interesting point to
    be noted  is the  Brillouin zone  in the  parameter space  for the
    transformed  Hamiltonian  with  the  eigen  basis  $|u_{nq}\rangle
    $.  The states  $|\psi_n(q)\rangle$  and $|\psi_{n}(q+h)\rangle  $
    satisfy  the same  boundary condition  as that  of the  torus. The
    crystal momentum q is found to vary and the Bloch state picks up a
    Berry phase.  This is nothing but the Zak phase $$\gamma_{n}=\oint
    dq.\langle u_n(q)|i\Delta_q|u_n(q)\rangle.$$ For  the Kitaev chain
    (Hamiltonian $\mathcal{H}^{(1)}(k)$) the physics is all right, but in the presence of
    $\alpha k$ term         the        wavefunction$$|\psi_n(q)\rangle\neq
    |\psi_n(q+h)\rangle ,$$ and the parameter  of the Zak phase is not
    valid.

\end{document}